\documentclass[sigconf]{acmart}

\usepackage[utf8]{inputenc} 
\usepackage[T1]{fontenc}    
\usepackage{hyperref}       
\usepackage{url}            
\usepackage{booktabs}       
\usepackage{amsfonts}       
\usepackage{nicefrac}       
\usepackage{microtype}      
\usepackage{xcolor}         
\usepackage{enumitem}

\usepackage{graphicx}
\usepackage{amsmath}

\usepackage{tabularx}
\usepackage{listings}
\usepackage{xcolor}
\usepackage{graphicx}
\usepackage{subfigure}
\usepackage{booktabs} 
\usepackage{caption}
\usepackage{subfigure}
\usepackage{epsfig}
\usepackage{algorithm}
\usepackage{algorithmic}
\usepackage{hyperref}
\AtBeginDocument{%
  \providecommand\BibTeX{{%
    \normalfont B\kern-0.5em{\scshape i\kern-0.25em b}\kern-0.8em\TeX}}}

\newcommand\blfootnote[1]{%
  \begingroup
  \renewcommand\thefootnote{}\footnote{#1}%
  \addtocounter{footnote}{-1}%
  \endgroup
}

\copyrightyear{2022} 
\acmYear{2022} 
\setcopyright{acmcopyright}\acmConference[CIKM '22]{Proceedings of the 31st ACM
International Conference on Information and Knowledge Management}{October
17--21, 2022}{Atlanta, GA, USA}
\acmBooktitle{Proceedings of the 31st ACM International Conference on Information
and Knowledge Management (CIKM '22), October 17--21, 2022, Atlanta, GA, USA}
\acmPrice{15.00}
\acmDOI{10.1145/3511808.3557238}
\acmISBN{978-1-4503-9236-5/22/10}

\begin{document}

\author{Zihan Liu}
\orcid{0000-0001-6224-3823}
\affiliation{%
  \institution{$^1$ Zhejiang University}
  \city{Hangzhou}
  \state{Zhejiang}
  \country{China}
}
\affiliation{%
  \institution{$^2$ AI Lab, Westlake University}
  \city{Hangzhou}
  \state{Zhejiang}
  \country{China}
}
\email{liuzihan@westlake.edu.cn}

\author{Yun Luo}
\affiliation{%
  \institution{ Westlake University}
  \city{Hangzhou}
  \state{Zhejiang}
  \country{China}
}

\author{Lirong Wu}
\affiliation{%
  \institution{AI Lab, Westlake University}
  \city{Hangzhou}
  \state{Zhejiang}
  \country{China}}

\author{Siyuan Li}
\affiliation{%
  \institution{AI Lab, Westlake University}
  \city{Hangzhou}
  \state{Zhejiang}
  \country{China}}

\author{Zicheng Liu}
\affiliation{%
  \institution{AI Lab, Westlake University}
  \city{Hangzhou}
  \state{Zhejiang}
  \country{China}}

\author{Stan Z. Li}
\authornote{Corresponding author}
\affiliation{%
  \institution{AI Lab, Westlake University}
  \city{Hangzhou}
  \state{Zhejiang}
  \country{China}}
\email{stan.zq.li@westlake.edu.cn}

\renewcommand{\shortauthors}{Zihan Liu et al.}




\title{Are Gradients on Graph Structure Reliable in Gray-box Attacks?}


\begin{abstract}

Graph edge perturbations are dedicated to damaging the prediction of graph neural networks by modifying the graph structure. Previous gray-box attackers employ gradients from the surrogate model to locate the vulnerable edges to perturb the graph structure. However, unreliability exists in gradients on graph structures, which is rarely studied by previous works. In this paper, we discuss and analyze the errors caused by the unreliability of the structural gradients. These errors arise from rough gradient usage due to the discreteness of the graph structure and from the unreliability in the meta-gradient on the graph structure. In order to address these problems, we propose a novel attack model with methods to reduce the errors inside the structural gradients. We propose edge discrete sampling to select the edge perturbations associated with hierarchical candidate selection to ensure computational efficiency. In addition, semantic invariance and momentum gradient ensemble are proposed to address the gradient fluctuation on semantic-augmented graphs and the instability of the surrogate model. Experiments are conducted in untargeted gray-box poisoning scenarios and demonstrate the improvement in the performance of our approach.
\end{abstract}



\maketitle

\section{Introduction}

Graph Neural Networks (GNNs) demonstrate excellent performance on various applications with structural, relational data \cite{zhou2020graph}, such as traffic \cite{wang2020traffic}, recommendation systems \cite{guo2020survey}, and social networks \cite{wang2019mcne}.
As the prospects for the applications of GNNs expand rapidly, their reliability and robustness are beginning to be of interest.
Several works have presented experimental evidence that GNNs are vulnerable to adversarial attacks \cite{zugner2018adversarial,zugner2019adversarial,dai2018adversarial}.
They design undetectable perturbations which successfully mislead GNNs' prediction of targeted nodes or degrade the performance on a global scope.
Subsequently, many works have been carried out around the graph adversarial attack and defense \cite{xu2020adversarial,sun2018adversarial}.

Gray-box attacks allow attackers to access the training labels from the victim model.
The attacker aims to damage the prediction of the victim model by injecting indistinguishable perturbations into the graph.
The attacker should search for vulnerable edges and attack them by modifying the graph.
In the field of adversarial attack, gradients are widely-used for attacking attributes that are spatially continuous \cite{xu2020adversarial}.
However, for graphs, the sparseness and discreteness of the graph structure make it challenging to perturb the graph structure in the way of Fast Gradient Sign Method (FGSM) \cite{goodfellow2014explaining} or Projected Gradient Descent (PGD) \cite{madry2017towards}.
To solve this problem, Zugner et al. \cite{zugner2019adversarial} firstly introduce the meta-gradient on the graph structure to determine the perturbation.
The attacker chooses one edge at a time to perturb based on the saliency of the gradient and iterates this step until the entire attack budget is consumed.
Subsequent works focus on improving the attack strategy after deriving the gradient and the surrogate model \cite{lin2020exploratory,liu2022surrogate}. However, few works focus on whether the saliency of the gradient on the graph structure is reliable.

Meta-gradients are demonstrated to be noisy in attribution problems \cite{sundararajan2017axiomatic}.
The gradients on the graph structure originate from the aggregation of node features, which means noises are equally propagated into the structural gradients.
Moreover, edge flipping changes the value of edges across a considerable step size (i.e., adding edges are from 0 to 1 and deleting edges are from 1 to 0).
Existing methods select the edge to be flipped based on the saliency of the gradient.
It is worth noting that, during the edge flipping, the structural gradient varies since the aggregation of node features is influenced by the edge value.
We consider edge perturbation to be a problem of modeling continuous gradient distributions rather than a discrete problem.
However, the structural gradients are imprecisely assumed to be constant, ignoring the variance in the edge-flip interval.\blfootnote{$^\dag$Corresponding author: Stan Z. Li.}

This paper points out the gradient errors that negatively affect the untargeted gray-box attacks on graph structure.
The discreteness of edges leads previous works to consider edge perturbation as a discrete problem about gradients.
However, the gradient at the current edge state commonly used in previous works is inaccurate for describing the gradient over the edge-flip interval.
We propose to transform the edge perturbation from a discrete problem to an approximation of a continuous problem and propose \textit{edge discrete sampling} to reduce this error.
Edge discrete sampling calculates the gradient of the transition process between the edge-flip interval in batches.
It reduces the error from the discrete approximation to the continuous gradient distribution, which performs a more accurate structural gradient estimation.
Since the space of edge perturbations is about the square of the number of nodes, the computational cost is unacceptable if the edge discrete sampling is performed for the whole space.
\textit{Hierarchical candidate selection} is then proposed to reduce the computational complexity.
It retains a bag of candidate edges that are more likely to be effective than processing the whole space.
In this step, we exploit the saliency of the gradient on the graph structure while the error still exists.
The random initialization of parameters leads to variance in surrogate model training, which affects the structural gradients via back-propagation.
Besides, the error occurs in gradient fluctuations on semantically identical graph augmentations.
These two errors are to be considered since gray-box attacks focus on attack transferability, so the gradient information from the surrogate model is expected to be more general and representative.
To address these errors, we first propose \textit{momentum gradient ensemble} to mitigate the instability of the structural gradients provided by the surrogate model at each attack iteration.
Then, we propose surrogate \textit{semantic invariance} based on graph augmentation in a limited semantic range.
These two methods of reducing the error on the gradient of the graph structure allow hierarchical candidate selection to provide better quality edge candidates.
Candidate selection ensures that the computational cost of edge discrete sampling does not grow exponentially, allowing the attacker to perform poisoning attacks.

The contributions of this paper are summarized as follows: 
\vspace{-1pt}
\begin{itemize}[leftmargin=0.6cm]\setlength{\itemsep}{-1pt}
\item We analyze the errors in structural gradients caused by model instability and the discreteness of graph structure in untargeted gray-box attacks.
\item We propose edge discrete sampling to approximate the continuous distributions of gradient over the edge-flip interval.
Hierarchical candidate selection is proposed to ensure that the computational complexity does not explode.
\item We propose semantic invariance and momentum gradient ensemble to address the gradient fluctuation on semantic graph augmentation and the instability of the surrogate model.
\item We demonstrate the improvement of our approach and prove each module's effectiveness in the ablation study.
\end{itemize}

\section{Preliminaries}

Before presenting the methodology and demonstrating the experiments, we first introduce the notations and backgrounds in this section. 
The notations used in the following sections can be referenced in Subsection 2.1. 
Subsection 2.2 introduces how to obtain the gradient on the graph structures by attacking loss in a generic attack strategy with edge perturbations.

\subsection{Notations}

A graph $G$ is represented as $G = (V,E,X)$, where $V=\{v_1,v_2,... ,v_n\}$ is the node set, $E \subseteq {V}\times{V}$ is the edge set, and $X$ is the attribution set.
In node classification tasks, a vertex $v_i$ is a sample with its features $x_i\in{\mathbb{R}^{d}}$ and label $y_i\in{\mathbb{R}^{c}}$.
A total of $N$ samples in the graph are distributed across $c$ classes.
The edges are represented as a binary adjacent matrix $A=\{0,1\}^{N\times N}$, where $A_{ij}=1$ if $(i,j)\subseteq E$.
We can also represent a graph with an adjacent matrix and nodes' attributes $G=(A,X)$.
For a gray-box attack, the attacker constructs a surrogate model, denoted by $f_{\theta}$, to simulate the process of the victim model being attacked.
The prediction of $f_{\theta}$ is denoted by the probability distribution $P_{v_i}$.
We use $\mathcal{L}$ to represent the loss functions.
The perturbed graph is indicated as $G'=(A',X')$ to distinguish the perturbed graph from the original graph.
The attack is limited by a budget $\Delta$.

\subsection{Edge Perturbations}

In the case of untargeted edge perturbation, the attacker is restricted to perturbing only by modifying the adjacency matrix (i.e., flipping edges).
The $l_0$ norm of the changes in the perturbed adjacency matrix with respect to the original one is bounded by the attacker's budget $\Delta$.
For an undirected graph, $\Delta$ is set as:
\begin{equation}
\lVert A-A'\rVert_0 \leq 2\Delta,
\end{equation}
where the budget $\Delta$ is generally equal to or less than 5\% of the number of edges in the original graph.

Gradient-based attack models now become mainstream methods of edge perturbations on the graph structure \cite{zugner2019adversarial,xu2019topology,lin2020exploratory}.
In contrast to the gradient-based attacks widely used in computer vision \cite{yuan2019adversarial}, the discrete graph structure restricts the gradient from being added directly to the adjacency matrix.
The gradient on the adjacent matrix (i.e. graph structure) \cite{zugner2019adversarial} $A^{grad}$ can be derived by the following equations:
\begin{equation}\label{surrogate training}
    {\theta}^*=\underset{\theta}{\mathrm{argmin}}\, \mathcal{L}_{ce}(f_\theta(G),Y),
\end{equation}
\begin{equation}\label{gradient deriving}
    A^{\,grad}=\nabla_{A}\mathcal{L}_{atk}(f_{\theta^*}(G)),
\end{equation}
where $\mathcal{L}_{atk}$ is the attack loss function and $f_{\theta^*}$ is the properly trained surrogate model.
To facilitate understanding, we elaborate in a more intuitive way of deriving $A^{\,grad}$. 
First, a GNN surrogate model $f_{\theta^*}$ is trained until it fits the training samples. Subsequently, the attack loss is backpropagated through the surrogate model generating gradients on the input adjacency matrix.
The attack loss is expressed as $\mathcal{L}_{atk}=-\mathcal{L}_{CE}$, which is a negative cross-entropy loss.
For the edge between nodes $v_i$ and $v_j$, if $A_{i,j}=1$ and $A^{grad}_{i,j}<0$, or if $A_{i,j}=0$ and $A^{grad}_{i,j}>0$, then flipping edge $E_{i,j}$ is considered as a perturbation that has the potential to negatively affect the victim model.
Among these edges, the one with the most significant gradient value is considered the most desirable perturbation for the current graph.
The process of perturbing the graph using the gradient information can be represented as:
\begin{equation}\label{attack}
    G'_{t}=\phi(\nabla_{A}\mathcal{L}_{atk}(f_{\theta^*}(G'_{t-1})),A'_{t-1}),
\end{equation}
where $\phi$ denotes the strategy for choosing the edge to be attacked.
The factors that influence the perturbation include the attack loss $\mathcal{L}_{atk}$ as well as the strategy $\phi$ and the surrogate model $f_{\theta^*}$.

\section{Methodology}

This section introduces the errors of the gradient on the graph structure and the methods to solve these errors. 
Section 3.1 first analyzes the error caused by interpreting edge perturbations as a discrete problem and proposes the solution \textit{edge discrete sampling}.
To rationalize the computational cost, we propose \textit{hierarchical candidate selection}.
It selects a bag of edge candidates based on the meta-gradient on the graph structure so that edge discrete sampling requires only a small number of edges to be processed in batches. 
When we rethink the meta-gradient on the graph structure, we find room for improving the reliability of the structural gradient. 
We analyze and discuss this part in Sections 3.2 and 3.3 and give the corresponding solutions.
Section 3.4 describes the overall attack flow.

\subsection{Error Caused by Edges' Discreteness}

\begin{figure}[t]
    \centering
    \includegraphics[width = 0.4\textwidth]{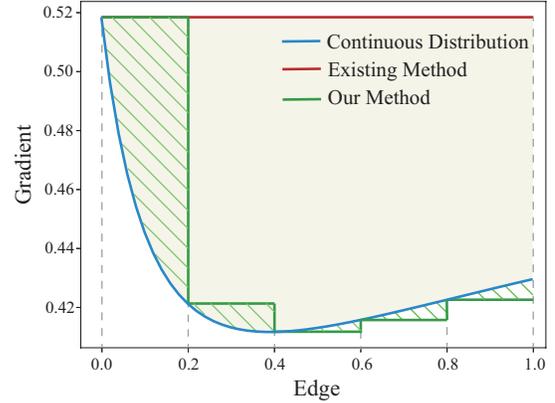}
    \caption{Illustration of the error generated by our method and the existing method in estimating gradient distribution on edge. The blue curve shows the continuous distribution. The thick lines in green and red are the estimation of our method and the existing method, while their estimation error to the continuous distribution is presented in thin green slash and light grey background.}
    \label{fig:3}
\end{figure}

As indicated by Eq.\ref{surrogate training}\&\ref{gradient deriving}, the gradient on $A_{ij}$ is the partial derivative of the attack loss $\mathcal{L}_{atk}$ to the adjacent matrix. 
In the existing approaches, the attackers treat the gradients on graph structure as a discrete problem of choosing the perturbations directly based on the saliency of the gradient \cite{zugner2019adversarial,lin2020exploratory}.
This means that the previous approach assumes that the gradient value is maintained at its value on the original state of the edge during the edge flipping (i.e., the state of an edge turns from 0 to 1 or from 1 to 0).
We give an example of the gradient approximation error introduced by this approach in Fig.\ref{fig:3}.
In contrast, we consider the gradient used to determine the edge perturbation as a continuous problem with continuous distribution approximation.

In Fig.\ref{fig:3}, the blue curve indicates the gradient distribution with edge value between 0 and 1, and the red line indicates the approximation of this distribution by the existing approaches. 
The light yellow shading indicates the error of the previous method compared to the ground truth continuous gradient distribution. Assuming that the interval for deriving the continuous distribution shown in the blue curve is $\delta$, the time complexity required to compute the gradient distribution for each edge is $O(N^2/\delta)$.
The problem is that the computational cost of calculating the gradient distribution for all edges is unacceptable.
In order to reduce the computational complexity, we propose hierarchical candidate selection. 
As mentioned in Section 2.2, the gradients on the adjacency matrix $A^{\,grad}$ can be derived from a trained surrogate model by backpropagation of the attack loss.
The first candidate selection step removes edges (set the saliency of the gradient to 0) where the sign of the gradient and the change of the edge does not match \cite{zugner2019adversarial}, after which the gradient matrix is processed to be $\Tilde{A}^{grad}$.
For the second step, we sort the rest of candidates according to the saliency of the gradient values in descending and retain the top $C$ edges as candidates: 
\begin{equation}\label{top}
    S=\{e_{(u_1,v_1)},...,e_{(u_c,v_c)}\} = \underset{e=(u,v)}{top_{\,C}}(\Tilde{A}^{grad}),
\end{equation}
where function $top_{\,C}$ extracts $C$ edge candidates of high saliency from the gradient matrix.
For these candidates, we reduce the time complexity to $O(C/\delta)$, where $C << N^2$. Up to this point, the time complexity of the algorithm is still enormous because $\delta$ is tiny. 
Therefore, in order to reduce the error while being able to ensure computational efficiency, we propose a discrete sampling method to approximate the continuous gradient distribution. The expression of the this approximation is:
\begin{equation}\label{gradient3}
\begin{aligned}
    g_{uv}^{int} &= \int_{0}^{1} [\nabla_{A}\mathcal{L}_{atk}(f_{\theta^*}(A,X))]_{u,v} \, dA_{uv} \\ &\approx
    \lambda\sum_{s=1}^{1/\lambda}{[\nabla_{A}\mathcal{L}_{atk}(f_{\theta^*}(A_{uv}=s\lambda,X))]_{u,v}},
\end{aligned}
\end{equation}
where $g_{uv}^{int}$ represents the integral gradient as the edge flips from 0 to 1 and $f_{\theta^*}(A_{uv}=s\lambda,X)$ is the result of modifying $A_{uv}$ to a transitional value $s\lambda$ without retraining.
The solid thick green line in Fig.\ref{fig:3} indicates the approximation by our algorithm, and the area in the thin green slash indicates the error caused by ours.
Compared with the error of the previous method indicated by the blue shading, our method substantially reduces the error generated in the edge-flipping process. Based on the above algorithm, the time complexity of our method decreases to $O(C/\lambda)$.
Considering that a perturbation can add or delete an edge, we adopt the one with the highest saliency of integrated gradient as a perturbation:
\begin{equation}\label{xxx}
    e'=\underset{e=(u,v)}{top_{\,1}}(\{(1-2A_{uv})\,g_{uv}^{int} \,|\, (u,v)\in S\}),
\end{equation}
where $e'$ denotes the selected edge to be perturbed, $(1-2A_{uv})$ is to invert the value of $g_{uv}^{int}$ for those candidates to delete an edge.

To further increase the computational efficiency, we introduce batch processing of candidates. 
A batch contains batch size $bs$ candidates. 
Since the gradients on the adjacent matrix come from the aggregation of node features, for a general 2-layer graph neural network, the state of one edge will influence the gradient on the other when the same node joints two edges. 
In other words, simultaneously changing the states of both edges causes a small amount of error.
However, for a batch of candidates selected from $N^2$ space, the probability that the candidates happen to be connected is minimal. 
Therefore we adopt batch processing which reduces the time complexity to $O(C/(\lambda*bs))$.

The method mentioned in Section 3.1 has \textit{a high dependence on the gradients on the adjacency matrix} $A^{\,grad}$.
Hierarchical candidate selection selects $C$ from the $N^2$ space, which requires strong reliability and stability of the gradients.
In Sections 3.2 and 3.3, we discuss the errors present in $A^{\,grad}$ as well as their solutions.

\subsection{Error Caused by Uncertainty of Model Optimization}

\begin{figure}[t]
    \centering
    \includegraphics[width = 0.40  \textwidth]{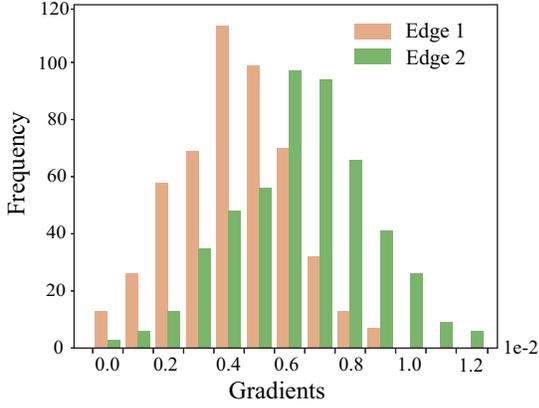}
    \caption{Possibility distributions of the gradient on two example edges. The x-axis is the interval of the gradient values, and the y-axis is the frequency of the gradient in the interval over 500 replicate experiments.}
    \label{fig:2}
\end{figure}

When the samples are not dense enough to describe the manifold of the data, the model is prone to fall into local optima.
The local optima of a model based on backpropagation optimization is related to the initialization of the mapping function of the neural network (i.e., the initialization of the learnable weights).
For the surrogate model, it tends to perform differently after retraining with different parameter initializations (i.e., in Equation \ref{surrogate training} different $\theta$ leads to different $\theta^*$).
We give an example to verify the uncertainty of structural gradients $A^{\,grad}$, shown in Fig.\ref{fig:2}.

Fig.\ref{fig:2} shows the gradients on two edges of the consistent graph after retraining the surrogate model with different weight initializations.
We can see that the gradient expectation of Edge 2 is around 0.007, which is higher than 0.004 of Edge 1. 
We conduct experiments showing that attacking Edge 2 is more effective than attacking Edge 1.
However, due to the randomness of weight initialization, the gradient of Edge 1 is possible to be higher than that of Edge 2, thus misleading the attacker to make a wrong judgment.

To enhance the reliability of $A^{\,grad}$ at each attack iteration, we expect each structural gradient obtained from the surrogate model to be as close as possible to its expectation in such distribution shown in Fig.\ref{fig:2}.
To this end, there is an easy way to solve this problem using an ensemble algorithm. 
We sample several weight initializations and integrate the gradients from retrained surrogate models to approximate the expectations of the structural gradients, reducing the probability of the gradient being sampled to outliers. 
The expression of the gradient ensemble is:
\begin{equation}\label{gradient2}
    A^{\,grad}=\frac{1}{k}\sum_{i=1}^k{A^{\,grad}_{\,\theta^*_i}},
\end{equation}
where $\theta^*_i$ represents the parameters of the surrogate model after training under initialization $\theta_i$ and $k$ denotes the number of integrations. 
This is an ensemble-based method that randomly initializes the model parameters with the identical uniform distribution between $[0, 1]$.
However, retraining the surrogate model is unacceptable, for it causes $k$ times the computational cost to be spent.

Considering that the attack on graph edges is a perturbation-by-perturbation iterative process, we propose a momentum gradient ensemble as a more efficient solution.
The variation between perturbed graphs $G^{'}_{t}$ and $G^{'}_{t-1}$ is limited.
The structural gradient $A^{\,grad}_t$ at iteration $t$ can reduce the instability of the gradient brought by each retraining of the surrogate model by fusing the structural gradient at previous iterations.
The derivation of the structural gradient at attack iteration $t$ is redefined as:
\begin{equation}\label{momentum}
    A^{\,grad}_t= \nabla_{A}\mathcal{L}_{atk}(f_{\theta^*}(G^{'}_{t})) + pA^{\,grad}_{t-1},
\end{equation}
where $p$ is the coefficient of the momentum term.
Compared with the ensemble method in Equation \ref{gradient2}, our momentum-based method consumes no additional computational cost. 
It makes full use of the structural gradients from previous iterations and avoids retraining the surrogate model multiple times in a single iteration.

\subsection{Error Caused by Model's Unrobustness}

The gradients of continuous data features are demonstrated to be noisy \cite{sundararajan2017axiomatic}.
Similar to data features, the graph structure is explicitly involved in the forward process in GNNs.
Taking a 2-layer GCN as an example, the expression of the forward process of graph neural network and the structural gradient is shown below.
\begin{equation}\label{5}
    f_{\theta}(A,X)=softmax(\hat{A}\tau{(\hat{A}XW^{(0)})}W^{(1)})
\end{equation}
\begin{equation}\label{6}
    \mathcal{L}_{atk} = -\mathcal{L}_{CE} = \log {P(y_i|f_{\theta}(A,X))},
\end{equation}
\begin{equation}\label{7}
    \nabla_{A_{ij}}\mathcal{L}_{atk} = \frac{\partial \log {P(y_i|f_{\theta})(A,X)}}{\partial A_{ij}},
\end{equation}

where $\hat{A}$ is the normalized adjacent matrix, $\tau$ is the activation function, $W$ is the weight matrix, $\mathcal{L}_{atk}$ equals to $-\mathcal{L}_{ce}$ and $P(y'_i|f_{\theta})$ represents the prediction at label class $y_i$.
It can be seen from Eq.\ref{5},\ref{6}\&\ref{7} that the derivation of the structural gradient involves the features of the nodes (which is related to the message passing mechanism in GNNs), resulting in the noise on the features being able to be propagated to the structural gradient.
The graph structure explicitly participates in the forward process of the model, so both input features and model parameters are influencing factors of the structural gradient. 
Therefore, similar to the sample features, the structural gradient is noisy.

\begin{figure}
    \centering
    \includegraphics[width = 0.42\textwidth]{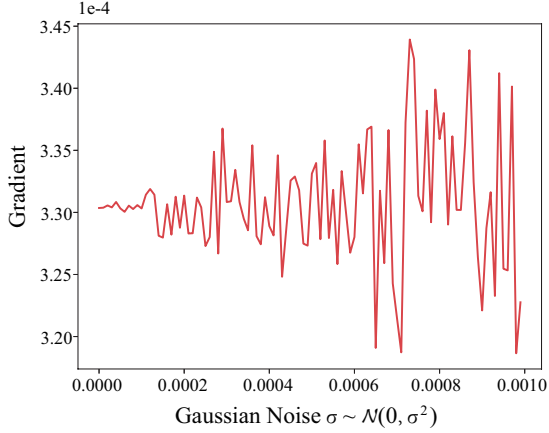}
    \caption{Illustration of the noisy gradients on graph structures. The curve represents the gradient change on an example edge under Gaussian noise on node attributes with variances $\sigma$.}
    \label{fig:1}
\end{figure}

Fig.\ref{fig:1} is an example which shows the gradients on an example edge on Citeseer \cite{sen2008collective}.
It is a citation network in which $10^{-3}$ can be considered as a non-disturbing semantic perturbation for node attributes. 
We inject Gaussian noise on the attributes of nodes $v_i$ and $v_j$ and their 1-hop neighbors, where the horizontal axis of Fig.\ref{fig:1} is the expectation of Gaussian noise and the vertical axis is the value of the gradient.
As can be seen in Fig.\ref{fig:1}, Gaussian noise with a standard deviation of $10^{-3}$ can make the gradient noisy. 
Considering the need for transferability for gray-box attacks, the instability of the surrogate model in the semantic range will affect the attacker's estimation of the retrained victim model.

It is worth noting that the noise is added to the node attributes rather than the graph structure. 
We consider adding noise to the graph structure would harm the graph homophily. 
For multi-class tasks, the majority of the noises are added between inter-class edges while the minority are added between intra-class edges.
This leads to the fact that adding Gaussian noise to the graph structure means \textit{decreasing the homophily ratio of the graph}.
Therefore, adding Gaussian noise on the node attributes is relatively reasonable than the graph structure.

In order to maintain the consistency of gradients in semantic graph augmentations, we propose a semantic invariance strategy based on graph augmentation.
Computing the expectation of gradient over a high-dimensional augmentation space is intractable, so we approximate this expectation by averaging sampled augmentations in the local space of the original graph. The semantic invariance is expressed mathematically as:
\begin{equation}\label{gradient1}
    A^{\,grad}_{si}=\frac{1}{n}\sum_{i=1}^n{\nabla_{A}\mathcal{L}_{atk}(f_{\theta^*}(A,X+\mathcal{N}(0,\sigma^2)))},
\end{equation}
where $n$ is the number of samples, and $\mathcal{N}(0, \sigma^2)$ represents a matrix of Gaussian noise with variable standard deviation $\sigma$.
This method preserves the invariance of gradients on semantically consistent augmented graphs.
The variance of Gaussian noise is a hyperparameter specific to the dataset. 
We empirically select the hyperparameter with the optimal performance on the validation set for testing based on grid search. 
For example, on Citeseer, the variance of the Gaussian noise is set to 5e-4. 
We verify that Gaussian noise at this variance has little effect on the classification accuracy of GNN, and that is how we define the noise as semantically non-disturbed.

\subsection{The Overall Attack Model}

This section describes the implementation of the above three error reduction methods in the attack model. 
Algorithm \ref{alg} is used to illustrate the whole flow of our proposed attack.

\begin{algorithm}[h]
\caption{Attack Pipeline}
\label{alg}
\begin{algorithmic}[1]
\REQUIRE
Original graph $G=(A,X)$, label of training data $Y$, attack budget $\Delta$, momentum coefficient $p$;
\ENSURE
Graph with edge perturbations $G'=(A',X)$;
\STATE $t = 0; A'=A$;
\WHILE{$\lVert A-A'\rVert_0 < 2\Delta$}
\STATE Train surrogate model $f_{\theta^*}$ with $G'=(A',X)$ and $Y$;

\STATE Derive the semantic invariance structrual gradient $A^{\,grad}_{si}$;

\STATE stack the structural gradient with that at the previous iteration in the way of momentum $A^{\,grad}_{t} = A^{\,grad}_{si} + pA^{\,grad}_{t-1}$ ($A^{\,grad}_{0} = A^{\,grad}_{si}$ at $t=0$);
\STATE Filter the candidates whose gradient symbols and edge states mismatch, and take the absolute value to get the saliency $\Tilde{A}^{grad}_t$;
\STATE Retain top $C$ edges in $\Tilde{A}^{grad}_t$ as candidate set $S_t$ and $S_{t\cdot copy}$
\WHILE{$S_t$ is not none}
\STATE Sample a batch $b$ of edge candidates from $S_t$;
\STATE Derive the integral gradient $g^{int}_{uv}$ of each edge $e_{uv}$ in batch $b$ at the edge flip interval [0,1];
\STATE Remove the batch $b$ from $S_t$;
\ENDWHILE
\STATE Extract the edge $e_{mn}$ with the most significant $g^{int}_{mn}$ in edge candidates $S_{t\cdot copy}$;
\STATE Update the perturbed graph $A'[m,n]=1-A'[m,n]$ and $A'[n,m]=1-A'[n,m]$;

\STATE $t = t + 1$;
\ENDWHILE
\end{algorithmic}
\end{algorithm}

The whole attack process is decomposed into $\Delta$ iterations, with one edge perturbed in each iteration.
At the beginning of each iteration, a surrogate model is trained using the perturbed graph $G'$ (Equation \ref{surrogate training}) to simulate the victim model under poisoning attacks.
With the trained surrogate model, we derive the semantic invariance structural gradient $A^{\,grad}_{si}$ following Equation \ref{gradient1}.
Then we add the momentum term onto the structural gradient to minimize the error arising from the model's local optima (Equation \ref{momentum}).
Line 6-7 in the Algorithm describe the hierarchical candidate selection, which is to choose candidate set $S_t$ from the $N^2$ space.
Line 8-11 in Algorithm perform edge discrete sampling (Equation \ref{gradient3}) to approximate the continuous gradient distribution for each edge in $S_t$.
Afterward, we select the edge with the most significant integral gradient $g^{int}_{mn}$ as the perturbed edge at the $t$-th iteration.
Finally, The perturbed graph structure is updated, and the $t+1$ iteration is started.

This algorithm ensures that the surrogate model needs to be trained only once in each iteration. 
Therefore, it achieves high computational efficiency while reducing the errors in structural gradients.
\begin{table*}[htp]
\begin{center}
\caption{Experimental results comparing our method with other methods. The group 'Clean' denotes the performance for unperturbed graph. The victim models are GraphSages and GCNs. The results are shown in the classification accuracy (\%) under perturbation rate 5\% on Cora, Cora-ML, Citeseer and Polblogs. The best results from experiments are bold.}
\label{table1}
\setlength\extrarowheight{1pt}

\begin{tabular}{ccccccccccc}
\hline
                              & \multicolumn{2}{c}{Cora}     & \multicolumn{2}{c}{Cora-ML}  & \multicolumn{2}{c}{Citeseer} & \multicolumn{2}{c}{Polblogs}  \\ \hline
Victim                     & GraphSage           & GCN          & GraphSage           & GCN          & GraphSage           & GCN          & GraphSage           & GCN   \\ \hline
Clean                        & 80.8$\pm$0.4 & 81.7$\pm$0.3    & 81.9$\pm$0.6 & 84.0$\pm$0.4   & 69.8$\pm$0.5 & 69.9$\pm$0.4     & 90.5$\pm$4.4 & 95.4$\pm$0.4      \\ \hline
Random                          & 80.7$\pm$0.6        & 81.2$\pm$0.3       & 80.6$\pm$0.8        & 82.8$\pm$0.4       & 68.7$\pm$0.7        & 67.8$\pm$0.4       & 78.1$\pm$1.1        & 84.0$\pm$0.9      \\
DICE                          & 80.0$\pm$0.4        & 80.9$\pm$0.5       & 80.7$\pm$0.9        & 82.5$\pm$0.4       & 69.1$\pm$0.8        & 69.0$\pm$0.4       & 76.1$\pm$1.9        & 79.2$\pm$0.7       \\
EpoAtk                        & 78.9$\pm$0.8   & 77.0$\pm$0.6  & 80.2$\pm$0.7   & 81.3$\pm$0.4  & 68.8$\pm$0.5   & 66.3$\pm$0.4  & 69.4$\pm$1.8   & 82.6$\pm$0.3   \\
Meta-Train                    & 76.1$\pm$1.0   & 76.4$\pm$0.2  & 77.7$\pm$1.8   & 79.0$\pm$0.3  & 67.9$\pm$0.9   & 65.5$\pm$0.4  & 71.2$\pm$4.0   & 83.5$\pm$0.7   \\
Meta-Self & 74.9$\pm$0.8 & 75.8$\pm$0.4  & 76.3$\pm$1.6   & 76.2$\pm$0.3   & 60.8$\pm$0.6  & 60.4$\pm$0.4   & \textbf{64.1$\pm$2.6}  & 75.8$\pm$0.5    \\ 
\textbf{AtkSE(ours)}                   & \textbf{73.3$\pm$0.6}    & \textbf{73.7$\pm$0.4}   & \textbf{75.4$\pm$1.0}    & \textbf{74.0$\pm$1.0}   & \textbf{60.7$\pm$0.4}    & \textbf{59.5$\pm$0.5}   & 65.6$\pm$0.8    & \textbf{71.5$\pm$0.3}   \\  \hline
\end{tabular}
\end{center}
\end{table*}

\section{Experiments}

We present experiments to demonstrate the effectiveness of our proposed attack model, named AtkSE \footnote{https://github.com/Zihan-Liu-00/AtkSE} (Attacking by Shrinking Errors).
The experimental settings are detailed in Section 4.1.
In the following two sections, we compare our approach with other gray-box untargeted poisoning attack methods and provide ablation studies to verify our proposed improvements' validity.
In Section 4.4, we provide the gradients' distribution on the edges' values between 0 and 1 to prove the necessity for our proposed approximation method for continuous gradient distribution.

\subsection{Experimental Settings}

\begin{table*}[htp]
\begin{center}
\caption{Ablation study of our proposed error reduction methods, where AtkSE-H ablate the hierarchical candidate selection as well as discrete sampling in Sec 3.1, AtkSE-M ablates the gradient ensemble in Sec 3.2, and AtkSE-S ablates the semantic invariance in Sec 3.3. The highest accuracy in the ablation is highlighted by underlining, representing the module that is ablated as more important in that experiment.}
\label{table2}
\setlength\extrarowheight{1pt}

\begin{tabular}{ccccccccccc}
\hline
                              & \multicolumn{2}{c}{Cora}     & \multicolumn{2}{c}{Cora-ML}  & \multicolumn{2}{c}{Citeseer} & \multicolumn{2}{c}{Polblogs}  \\ \hline
Victim                     & GraphSage           & GCN          & GraphSage           & GCN          & GraphSage           & GCN          & GraphSage           & GCN   \\ \hline
AtkSE                        & 73.3$\pm$0.6 & 73.7$\pm$0.4    & 75.4$\pm$1.0 & 74.0$\pm$1.0   & 60.7$\pm$0.4 & 59.5$\pm$0.5     & 65.6$\pm$0.8 & 71.5$\pm$0.3      \\ \hline

AtkSE-H                       & 73.9$\pm$0.6   & 74.6$\pm$0.3  & 75.2$\pm$1.1   & \underline{75.3$\pm$0.8}  & 61.0$\pm$0.6   & \underline{60.4$\pm$0.5}  & 67.6$\pm$0.9   & \underline{72.4$\pm$0.3}   \\
AtkSE-M                          & \underline{74.0$\pm$0.7}        & \underline{75.2$\pm$0.2}       & 75.5$\pm$1.0        & 74.7$\pm$0.7       & \underline{61.6$\pm$0.5}        & 60.3$\pm$0.5       & \underline{68.0$\pm$1.3}        & 71.8$\pm$0.1       \\
AtkSE-S                          & 73.8$\pm$0.6        & 74.5$\pm$0.4       & \underline{75.6$\pm$1.1}        & 75.1$\pm$0.7       & 60.6$\pm$0.8        & 60.0$\pm$0.4       & 66.1$\pm$2.3        & 72.0$\pm$0.3      \\

\hline
\end{tabular}

\end{center}
\end{table*}

\subsubsection{Datasets}
In this paper, we use the citation networks Citeseer \cite{sen2008collective}, Cora \cite{mccallum2000automating} and Cora-ML \cite{mccallum2000automating} as well as the social network Polblogs \cite{adamic2005political} as the datasets.
Consistent with the experimental setup of baselines, we randomly divide the dataset into 10\% of labeled nodes and 90\% of unlabeled nodes.
The labels of the unlabeled nodes are not provided to the attacker or the victim model, and they are only used when evaluating the performance of the victim model. 
\subsubsection{Victim Models}
The widely-used victim model is GCN \cite{kipf2016semi} used in baseline papers.
This paper extends GraphSage \cite{hamilton2017inductive} as a more advanced GNN victim model.
It is worth noting that there are some gray-box attacks, such as \cite{lin2020exploratory} and \cite{liu2022surrogate}, where the network architecture of the victim model is considered to be known.
The attack scenario in this paper considers that the victim model's architecture is unknown to study the poisoning attack's transferability better.
Therefore, the GCN victim model differs from the surrogate model in linearity and number of neurons.

We uniformly use a 5\% perturbation rate as the attack budget for imperceptibility needs.
We repeat the experiments ten times for each experimental scenario and present the mean and variance of each set of experiments in the results.
To ensure the fairness of the experiments, we test the perturbed graphs generated by each method with uniform and independent test files.
\subsubsection{Baselines}
Random, DICE \cite{waniek2018hiding}, EpoAtk \cite{lin2020exploratory}, Meta-Train \& Meta-Self \cite{zugner2019adversarial} are used as baselines in the experiments. 

\noindent\textbf{Random} removes or add edges randomly.

\noindent\textbf{DICE} randomly removes edges between nodes from the same class or adds edges between nodes from different classes.

\noindent\textbf{EpoAtk} is originally a white-box attack model, transferred to the gray-box attack in our experiments. It proposes an exploration strategy where the attacker chooses edges from a set of candidates.

\noindent\textbf{Meta-Self and Meta-Train} consider the adjacent matrix as hyper-parameters to optimize via meta-learning. Two attack models differ in the node subset used to calculate attack loss.

\subsubsection{Hyperparameters} In the implementation of our attack model, the interval of edge discrete sampling $\lambda$ is set to $0.2$, the number of edge candidates $C$ is set to $64$, the batch $bs$ is set to $16$, the momentum coefficient is set to $0.8$, and $n$ in semantic invariance is $5$. 

\subsection{Performance of AtkSE}

Table \ref{table1} shows the comparison of our approach with baselines on various datasets and victim models.
Among the eight experiments, our proposed AtkSE outperforms baselines in seven of them.
Meta-Self is the most effective baseline in other gradient-based baselines, followed by Meta-Train and EpoAtk. 
The randomness-based methods DICE and Random are the worst.
Our method ranks second behind the best effect in Polblogs/GraphSage.
When the victim models are GraphSages, our method improves the attack success rate over the second-best method by 1.6\%, 0.9\%, and 0.1\% on datasets Cora, Cora-ML, and Citeseer, respectively.
On Polblogs, our method ranks second, below the first place by 1.5\%.
This result may be due to the independence of victim models' representation learning and the difficulty of transferring attacks.
When the victim models are GCNs, our method is better than other methods across the board.
The attack performance of our method improves 2.1\%, 1.8\%, 0.9\%, and 4.3\% on Cora, Cora-ML, Citeseer, and Polblogs, compared to the second-place method Meta-Self.

Averaged over experiments, the attack effect of our method is 1.275\% higher than that of Meta-Self.
Our proposed AtkSE outperforms the second-ranked model by more than 1\% in four experiments and 2\% in two experiments.
Experiments prove that our attack model, AtkSE, is more effective than other methods. 
It proves that errors exist in the previous methods and that reducing these errors can improve the effectiveness of the attack model.

\subsection{Ablation Study}

To verify the effectiveness of each error reduction module, we conduct the ablation study.
We ablate the three error reduction modules in Sections 3.1 (\textbf{H}ierarchical candidate selection and edge discrete sampling), 3.2 (\textbf{M}omentum gradient ensemble), and 3.3 (\textbf{S}emantic invariance), respectively.
The ablated attack models are denoted as AtkSE-H, AtkSE-M, and AtkSE-S.
Table \ref{table2} shows the results of the ablation study.

In the ablation experiments, the worse the attack of the ablated model is, the more critical the ablated module is.
AtkSE-M has the best accuracy on Cora, Citeseer/GraphSage, and Polblogs/GraphSage.
It ranks second on Cora-ML/GraphSage and Citeseer/GCN and worse than the other scenarios.
AtkSE-H has the highest accuracy when the dataset is Cora-ML, Citeseer and Polblogs, and the victim model is GCN.
AtkSE-H ranks second on four experiments and worse on Cora-ML/GraphSage.
AtkSE-S ranks first on Cora-ML/GraphSage, while it ranks lower on five experiments.
AtkSE-M has the highest accuracy on four experiments, while AtkSE-H and AtkSE-S have three and merely one, respectively.

Overall, the momentum gradient ensemble is the module that enhances the attack model the most.
The results indicate that the modules are ranked from highest to lowest importance as the momentum gradient ensemble, the hierarchical candidate selection and edge discrete sampling, and the semantic invariance.
By comparing ablated models with AtkSE, the experiments prove the effectiveness of each module in AtkSE.

\subsection{Gradient between the Edge-flip Interval}

\begin{figure}[t]
    \centering
    \includegraphics[width = 0.44\textwidth]{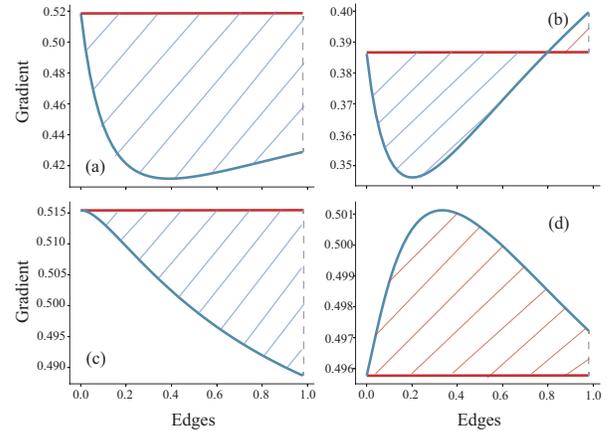}
    \caption{Illustrations of several variations of the structural gradient between the edge-flip interval.}
    \label{fig:visual}
\end{figure}

In this paper, graph edge perturbation is considered a problem of modeling a continuous distribution of gradients on edges.
A possible challenge for our approach is whether the problem is worthy of being solved as a continuous problem over the edge-flip interval.
To answer this question, we give examples of continuous gradient distributions of edge candidates in Fig.\ref{fig:visual}.
We can see that the gradients of the edges are continuous and smooth on the interval.
We use the blue slash to indicate cases where the estimate is above the actual distribution and the red slash to indicate cases where the estimate is below the actual distribution.
Referring to the error analysis in Section 3.1, in Fig.\ref{fig:visual}, the previous methods adopt a gradient value higher than the integral of the continuous distribution on the (a,b,c) plot and lower than the integral on the (d) plot.
Our approach considers the variation of gradients and transforms the edge perturbation from a discrete problem to a continuous gradients modeling problem.
The error observed in Fig.\ref{fig:visual} proves the necessity for modeling continuous structural gradients. 
It also demonstrates why our approach improves the effectiveness of the attack.

\section{Related Works}
\subsection{Graph Adversarial Attack}
Graph adversarial attacks aim to disrupt the performance of graph neural networks using imperceptible attacks.
There are three ways to attack graph-structured data: modifying the node features, modifying the graph structure, and injecting nodes \cite{jinwei2020kddsurvey,chen2020survey}.
Among the researches in this field, more studies focus on modifying the graph structure \cite{dai2018adversarial,waniek2018hiding}, and node injection \cite{sun2020adversarial}, due to the specificity of graph data compared to general data.
Among the general imperceptibility measures, attacks on graph structure are constrained by the L0 norm \cite{zugner2019adversarial}, and node injections are constrained by the number of manual nodes and their degrees \cite{sun2020adversarial}.
Depending on the attacker's knowledge, attacks are classified as white-box attacks, gray-box attacks, and black-box attacks.
White-box attacks \cite{xu2019topology,wu2019adversarial} open all the information of victim models.
Gray-box \cite{lin2020exploratory,liu2022surrogate} attacks open labels of training set. 
Black-box attacks \cite{xu2020adversarial,ma2019attacking} allow the attackers to query the predictions of the victim models.
If the victim model is retrained, the attack is referred to as a poisoning attack, otherwise, it is referred to as an evasion attack.
This paper studies the gray-box poisoning attacks.
We aim to transfer the attack from the surrogate model to an unknown victim model, also referred to as the study of attack transferability.

\subsection{Graph Edge Perturbation}
This paper investigates attackers which globally perturb the graph structure.
Mainstream approaches are based on the gradient derived from the loss function by backpropagation on the graph structure (or the adjacent matrix).
Metattack \cite{zugner2019adversarial} is the first gradient-based edge perturbation work on graph networks.
It is a global poison attack model with a gray-box setting and the basis of other gradient-based edge perturbation methods.
The attack strategy of Metattack is to search for the edge with the most significant gradient by the gradient on the adjacency matrix for modification, modifying one edge per iteration until an upper limit of the budget is reached.
Another work \cite{xu2019topology} makes use of PGD \cite{madry2017towards} to search for an optimal perturbation matrix $S \in  \{0,1\}^n $.
EpoAtk \cite{lin2020exploratory} tries to add exploration to the attack model.
EpoAtk has a small probability of not directly perturbing the edge with the most significant gradient, but instead of generating a perturbation from many candidate edges based on the loss values.
These three works ignore whether the gradients they use are reliable.
They only use the gradient derived from the current state of the graph and do not take into account the variation of the gradients on edges over the perturbation interval from 0 to 1.
A method of integrated gradient on edges \cite{wu2019adversarial} is proposed to solve this problem. 
It requires traversing the adjacency matrix to score all edges and multiple gradient calculations to score each edge.
It needs to traverse the adjacency matrix to score all edges, which has an exploded computational complexity $O(N^2)$.
Moreover, it requires a massive amount of computing resources to perform a single time gradient calculation on all edges, limiting it to be used only in evasion attacks of small graphs, not any of the other scenarios.
In addition, these efforts do not consider the gradient instability during the model optimization process.
The above problems are addressed in our proposed method.

\section{Conclusion}

This paper proposes that the gradient on the graph structure in graph adversarial attack is subject to errors.
This paper aims to illustrate these errors and propose corresponding modules to reduce them and implement them into our proposed attack model AtkSE.
This paper first analyzes the error caused by treating graph structure attacks as a discrete problem with respect to gradients.
We propose edge discrete sampling to reduce this error by approximating the continuous distribution of gradients over the edge flipping interval [0,1].
To ensure the computational cost of this step, we propose hierarchical candidate sampling based on the gradient saliency of the graph structure to select a small bag of edges as the candidates to be attacked rather than the entire $N^2$ space.
Subsequently, we discuss the errors present in the gradient of the graph structure, including GNNs' unrobustness on the semantic space of the graph and the instability of the GNNs' representations due to the randomness of the parameter initialization.
We propose semantic invariance and momentum gradient ensemble to solve these two errors, respectively.
We integrate the above error reduction modules and propose the attack model AtkSE.
In the experiments, we validate the effectiveness of our proposed method by comparing it with state-of-the-art baselines and showing the ablation study in untargeted poisoning gray-box attacks. 
The results demonstrate that our approach improves the attacker's performance, proving the reliability of our discussion on the error analysis and the effectiveness of our approach.

\section*{Acknowledgement}
This work is supported in part by Ministry of Science and Technology of the People´s Republic of China (No. 2021YFA1301603) and National Natural Science Foundation of China (No. U21A20427).

\bibliographystyle{ACM-Reference-Format}
\bibliography{ref}


\begin{thebibliography}{27}


\ifx \showCODEN    \undefined \def \showCODEN     #1{\unskip}     \fi
\ifx \showDOI      \undefined \def \showDOI       #1{#1}\fi
\ifx \showISBNx    \undefined \def \showISBNx     #1{\unskip}     \fi
\ifx \showISBNxiii \undefined \def \showISBNxiii  #1{\unskip}     \fi
\ifx \showISSN     \undefined \def \showISSN      #1{\unskip}     \fi
\ifx \showLCCN     \undefined \def \showLCCN      #1{\unskip}     \fi
\ifx \shownote     \undefined \def \shownote      #1{#1}          \fi
\ifx \showarticletitle \undefined \def \showarticletitle #1{#1}   \fi
\ifx \showURL      \undefined \def \showURL       {\relax}        \fi
\providecommand\bibfield[2]{#2}
\providecommand\bibinfo[2]{#2}
\providecommand\natexlab[1]{#1}
\providecommand\showeprint[2][]{arXiv:#2}

\bibitem[Adamic and Glance(2005)]%
        {adamic2005political}
\bibfield{author}{\bibinfo{person}{Lada~A Adamic} {and}
  \bibinfo{person}{Natalie Glance}.} \bibinfo{year}{2005}\natexlab{}.
\newblock \showarticletitle{The political blogosphere and the 2004 US election:
  divided they blog}. In \bibinfo{booktitle}{\emph{Proceedings of the 3rd
  international workshop on Link discovery}}. \bibinfo{pages}{36--43}.
\newblock


\bibitem[Chen et~al\mbox{.}(2020)]%
        {chen2020survey}
\bibfield{author}{\bibinfo{person}{Liang Chen}, \bibinfo{person}{Jintang Li},
  \bibinfo{person}{Jiaying Peng}, \bibinfo{person}{Tao Xie},
  \bibinfo{person}{Zengxu Cao}, \bibinfo{person}{Kun Xu},
  \bibinfo{person}{Xiangnan He}, {and} \bibinfo{person}{Zibin Zheng}.}
  \bibinfo{year}{2020}\natexlab{}.
\newblock \showarticletitle{A survey of adversarial learning on graphs}.
\newblock \bibinfo{journal}{\emph{arXiv preprint arXiv:2003.05730}}
  (\bibinfo{year}{2020}).
\newblock


\bibitem[Dai et~al\mbox{.}(2018)]%
        {dai2018adversarial}
\bibfield{author}{\bibinfo{person}{Hanjun Dai}, \bibinfo{person}{Hui Li},
  \bibinfo{person}{Tian Tian}, \bibinfo{person}{Xin Huang},
  \bibinfo{person}{Lin Wang}, \bibinfo{person}{Jun Zhu}, {and}
  \bibinfo{person}{Le Song}.} \bibinfo{year}{2018}\natexlab{}.
\newblock \showarticletitle{Adversarial attack on graph structured data}. In
  \bibinfo{booktitle}{\emph{International conference on machine learning}}.
  PMLR, \bibinfo{pages}{1115--1124}.
\newblock


\bibitem[Goodfellow et~al\mbox{.}(2015)]%
        {goodfellow2014explaining}
\bibfield{author}{\bibinfo{person}{Ian~J Goodfellow}, \bibinfo{person}{Jonathon
  Shlens}, {and} \bibinfo{person}{Christian Szegedy}.}
  \bibinfo{year}{2015}\natexlab{}.
\newblock \showarticletitle{EXPLAINING AND HARNESSING ADVERSARIAL EXAMPLES}.
\newblock \bibinfo{journal}{\emph{stat}}  \bibinfo{volume}{1050}
  (\bibinfo{year}{2015}), \bibinfo{pages}{20}.
\newblock


\bibitem[Guo et~al\mbox{.}(2020)]%
        {guo2020survey}
\bibfield{author}{\bibinfo{person}{Qingyu Guo}, \bibinfo{person}{Fuzhen
  Zhuang}, \bibinfo{person}{Chuan Qin}, \bibinfo{person}{Hengshu Zhu},
  \bibinfo{person}{Xing Xie}, \bibinfo{person}{Hui Xiong}, {and}
  \bibinfo{person}{Qing He}.} \bibinfo{year}{2020}\natexlab{}.
\newblock \showarticletitle{A survey on knowledge graph-based recommender
  systems}.
\newblock \bibinfo{journal}{\emph{IEEE Transactions on Knowledge and Data
  Engineering}} (\bibinfo{year}{2020}).
\newblock


\bibitem[Hamilton et~al\mbox{.}(2017)]%
        {hamilton2017inductive}
\bibfield{author}{\bibinfo{person}{William~L Hamilton}, \bibinfo{person}{Rex
  Ying}, {and} \bibinfo{person}{Jure Leskovec}.}
  \bibinfo{year}{2017}\natexlab{}.
\newblock \showarticletitle{Inductive representation learning on large graphs}.
  In \bibinfo{booktitle}{\emph{Proceedings of the 31st International Conference
  on Neural Information Processing Systems}}. \bibinfo{pages}{1025--1035}.
\newblock


\bibitem[Jin et~al\mbox{.}(2021)]%
        {jinwei2020kddsurvey}
\bibfield{author}{\bibinfo{person}{Wei Jin}, \bibinfo{person}{Yaxing Li},
  \bibinfo{person}{Han Xu}, \bibinfo{person}{Yiqi Wang},
  \bibinfo{person}{Shuiwang Ji}, \bibinfo{person}{Charu Aggarwal}, {and}
  \bibinfo{person}{Jiliang Tang}.} \bibinfo{year}{2021}\natexlab{}.
\newblock \showarticletitle{Adversarial Attacks and Defenses on Graphs}.
\newblock \bibinfo{journal}{\emph{SIGKDD Explor. Newsl.}}
  (\bibinfo{year}{2021}), \bibinfo{pages}{19–34}.
\newblock


\bibitem[Kipf and Welling(2017)]%
        {kipf2016semi}
\bibfield{author}{\bibinfo{person}{Thomas~N. Kipf} {and} \bibinfo{person}{Max
  Welling}.} \bibinfo{year}{2017}\natexlab{}.
\newblock \showarticletitle{Semi-Supervised Classification with Graph
  Convolutional Networks}. In \bibinfo{booktitle}{\emph{5th International
  Conference on Learning Representations, {ICLR} 2017, Toulon, France, April
  24-26, 2017, Conference Track Proceedings}}.
\newblock


\bibitem[Lin et~al\mbox{.}(2020)]%
        {lin2020exploratory}
\bibfield{author}{\bibinfo{person}{Xixun Lin}, \bibinfo{person}{Chuan Zhou},
  \bibinfo{person}{Hong Yang}, \bibinfo{person}{Jia Wu}, \bibinfo{person}{Haibo
  Wang}, \bibinfo{person}{Yanan Cao}, {and} \bibinfo{person}{Bin Wang}.}
  \bibinfo{year}{2020}\natexlab{}.
\newblock \showarticletitle{Exploratory Adversarial Attacks on Graph Neural
  Networks}. In \bibinfo{booktitle}{\emph{2020 IEEE International Conference on
  Data Mining (ICDM)}}. IEEE, \bibinfo{pages}{1136--1141}.
\newblock


\bibitem[Liu et~al\mbox{.}(2022)]%
        {liu2022surrogate}
\bibfield{author}{\bibinfo{person}{Zihan Liu}, \bibinfo{person}{Yun Luo},
  \bibinfo{person}{Zelin Zang}, {and} \bibinfo{person}{Stan~Z Li}.}
  \bibinfo{year}{2022}\natexlab{}.
\newblock \showarticletitle{Surrogate Representation Learning with Isometric
  Mapping for Gray-box Graph Adversarial Attacks}. In
  \bibinfo{booktitle}{\emph{Proceedings of the Fifteenth ACM International
  Conference on Web Search and Data Mining}}. \bibinfo{pages}{591--598}.
\newblock


\bibitem[Ma et~al\mbox{.}(2019)]%
        {ma2019attacking}
\bibfield{author}{\bibinfo{person}{Yao Ma}, \bibinfo{person}{Suhang Wang},
  \bibinfo{person}{Tyler Derr}, \bibinfo{person}{Lingfei Wu}, {and}
  \bibinfo{person}{Jiliang Tang}.} \bibinfo{year}{2019}\natexlab{}.
\newblock \showarticletitle{Attacking graph convolutional networks via
  rewiring}.
\newblock \bibinfo{journal}{\emph{arXiv preprint arXiv:1906.03750}}
  (\bibinfo{year}{2019}).
\newblock


\bibitem[Madry et~al\mbox{.}(2018)]%
        {madry2017towards}
\bibfield{author}{\bibinfo{person}{Aleksander Madry},
  \bibinfo{person}{Aleksandar Makelov}, \bibinfo{person}{Ludwig Schmidt},
  \bibinfo{person}{Dimitris Tsipras}, {and} \bibinfo{person}{Adrian Vladu}.}
  \bibinfo{year}{2018}\natexlab{}.
\newblock \showarticletitle{Towards Deep Learning Models Resistant to
  Adversarial Attacks}. In \bibinfo{booktitle}{\emph{International Conference
  on Learning Representations}}.
\newblock


\bibitem[McCallum et~al\mbox{.}(2000)]%
        {mccallum2000automating}
\bibfield{author}{\bibinfo{person}{Andrew~Kachites McCallum},
  \bibinfo{person}{Kamal Nigam}, \bibinfo{person}{Jason Rennie}, {and}
  \bibinfo{person}{Kristie Seymore}.} \bibinfo{year}{2000}\natexlab{}.
\newblock \showarticletitle{Automating the construction of internet portals
  with machine learning}.
\newblock \bibinfo{journal}{\emph{Information Retrieval}} \bibinfo{volume}{3},
  \bibinfo{number}{2} (\bibinfo{year}{2000}), \bibinfo{pages}{127--163}.
\newblock


\bibitem[Sen et~al\mbox{.}(2008)]%
        {sen2008collective}
\bibfield{author}{\bibinfo{person}{Prithviraj Sen}, \bibinfo{person}{Galileo
  Namata}, \bibinfo{person}{Mustafa Bilgic}, \bibinfo{person}{Lise Getoor},
  \bibinfo{person}{Brian Galligher}, {and} \bibinfo{person}{Tina Eliassi-Rad}.}
  \bibinfo{year}{2008}\natexlab{}.
\newblock \showarticletitle{Collective classification in network data}.
\newblock \bibinfo{journal}{\emph{AI magazine}} \bibinfo{volume}{29},
  \bibinfo{number}{3} (\bibinfo{year}{2008}), \bibinfo{pages}{93--93}.
\newblock


\bibitem[Sun et~al\mbox{.}(2018)]%
        {sun2018adversarial}
\bibfield{author}{\bibinfo{person}{Lichao Sun}, \bibinfo{person}{Yingtong Dou},
  \bibinfo{person}{Carl Yang}, \bibinfo{person}{Ji Wang},
  \bibinfo{person}{Philip~S. Yu}, \bibinfo{person}{Lifang He}, {and}
  \bibinfo{person}{Bo Li}.} \bibinfo{year}{2018}\natexlab{}.
\newblock \showarticletitle{Adversarial Attack and Defense on Graph Data: A
  Survey}.
\newblock \bibinfo{journal}{\emph{arXiv preprint arXiv:1812.10528}}
  (\bibinfo{year}{2018}).
\newblock


\bibitem[Sun et~al\mbox{.}(2020)]%
        {sun2020adversarial}
\bibfield{author}{\bibinfo{person}{Yiwei Sun}, \bibinfo{person}{Suhang Wang},
  \bibinfo{person}{Xianfeng Tang}, \bibinfo{person}{Tsung-Yu Hsieh}, {and}
  \bibinfo{person}{Vasant Honavar}.} \bibinfo{year}{2020}\natexlab{}.
\newblock \showarticletitle{Adversarial attacks on graph neural networks via
  node injections: A hierarchical reinforcement learning approach}. In
  \bibinfo{booktitle}{\emph{Proceedings of The Web Conference 2020}}.
  \bibinfo{pages}{673--683}.
\newblock


\bibitem[Sundararajan et~al\mbox{.}(2017)]%
        {sundararajan2017axiomatic}
\bibfield{author}{\bibinfo{person}{Mukund Sundararajan}, \bibinfo{person}{Ankur
  Taly}, {and} \bibinfo{person}{Qiqi Yan}.} \bibinfo{year}{2017}\natexlab{}.
\newblock \showarticletitle{Axiomatic attribution for deep networks}. In
  \bibinfo{booktitle}{\emph{International Conference on Machine Learning}}.
  PMLR, \bibinfo{pages}{3319--3328}.
\newblock


\bibitem[Wang et~al\mbox{.}(2019)]%
        {wang2019mcne}
\bibfield{author}{\bibinfo{person}{Hao Wang}, \bibinfo{person}{Tong Xu},
  \bibinfo{person}{Qi Liu}, \bibinfo{person}{Defu Lian},
  \bibinfo{person}{Enhong Chen}, \bibinfo{person}{Dongfang Du},
  \bibinfo{person}{Han Wu}, {and} \bibinfo{person}{Wen Su}.}
  \bibinfo{year}{2019}\natexlab{}.
\newblock \showarticletitle{MCNE: an end-to-end framework for learning multiple
  conditional network representations of social network}. In
  \bibinfo{booktitle}{\emph{Proceedings of the 25th ACM SIGKDD International
  Conference on Knowledge Discovery \& Data Mining}}.
  \bibinfo{pages}{1064--1072}.
\newblock


\bibitem[Wang et~al\mbox{.}(2020)]%
        {wang2020traffic}
\bibfield{author}{\bibinfo{person}{Xiaoyang Wang}, \bibinfo{person}{Yao Ma},
  \bibinfo{person}{Yiqi Wang}, \bibinfo{person}{Wei Jin}, \bibinfo{person}{Xin
  Wang}, \bibinfo{person}{Jiliang Tang}, \bibinfo{person}{Caiyan Jia}, {and}
  \bibinfo{person}{Jian Yu}.} \bibinfo{year}{2020}\natexlab{}.
\newblock \showarticletitle{Traffic flow prediction via spatial temporal graph
  neural network}. In \bibinfo{booktitle}{\emph{Proceedings of The Web
  Conference 2020}}. \bibinfo{pages}{1082--1092}.
\newblock


\bibitem[Waniek et~al\mbox{.}(2018)]%
        {waniek2018hiding}
\bibfield{author}{\bibinfo{person}{Marcin Waniek}, \bibinfo{person}{Tomasz~P
  Michalak}, \bibinfo{person}{Michael~J Wooldridge}, {and}
  \bibinfo{person}{Talal Rahwan}.} \bibinfo{year}{2018}\natexlab{}.
\newblock \showarticletitle{Hiding individuals and communities in a social
  network}.
\newblock \bibinfo{journal}{\emph{Nature Human Behaviour}} \bibinfo{volume}{2},
  \bibinfo{number}{2} (\bibinfo{year}{2018}), \bibinfo{pages}{139--147}.
\newblock


\bibitem[Wu et~al\mbox{.}(2019)]%
        {wu2019adversarial}
\bibfield{author}{\bibinfo{person}{Huijun Wu}, \bibinfo{person}{Chen Wang},
  \bibinfo{person}{Yuriy Tyshetskiy}, \bibinfo{person}{Andrew Docherty},
  \bibinfo{person}{Kai Lu}, {and} \bibinfo{person}{Liming Zhu}.}
  \bibinfo{year}{2019}\natexlab{}.
\newblock \showarticletitle{Adversarial examples on graph data: Deep insights
  into attack and defense}. In \bibinfo{booktitle}{\emph{International Joint
  Conference on Artificial Intelligence}}.
\newblock


\bibitem[Xu et~al\mbox{.}(2020)]%
        {xu2020adversarial}
\bibfield{author}{\bibinfo{person}{Han Xu}, \bibinfo{person}{Yao Ma},
  \bibinfo{person}{Hao-Chen Liu}, \bibinfo{person}{Debayan Deb},
  \bibinfo{person}{Hui Liu}, \bibinfo{person}{Ji-Liang Tang}, {and}
  \bibinfo{person}{Anil~K Jain}.} \bibinfo{year}{2020}\natexlab{}.
\newblock \showarticletitle{Adversarial attacks and defenses in images, graphs
  and text: A review}.
\newblock \bibinfo{journal}{\emph{International Journal of Automation and
  Computing}} \bibinfo{volume}{17}, \bibinfo{number}{2} (\bibinfo{year}{2020}),
  \bibinfo{pages}{151--178}.
\newblock


\bibitem[Xu et~al\mbox{.}(2019)]%
        {xu2019topology}
\bibfield{author}{\bibinfo{person}{Kaidi Xu}, \bibinfo{person}{Hongge Chen},
  \bibinfo{person}{Sijia Liu}, \bibinfo{person}{Pin~Yu Chen},
  \bibinfo{person}{Tsui~Wei Weng}, \bibinfo{person}{Mingyi Hong}, {and}
  \bibinfo{person}{Xue Lin}.} \bibinfo{year}{2019}\natexlab{}.
\newblock \showarticletitle{Topology attack and defense for graph neural
  networks: An optimization perspective}. In \bibinfo{booktitle}{\emph{28th
  International Joint Conference on Artificial Intelligence, IJCAI 2019}}.
  \bibinfo{pages}{3961--3967}.
\newblock


\bibitem[Yuan et~al\mbox{.}(2019)]%
        {yuan2019adversarial}
\bibfield{author}{\bibinfo{person}{Xiaoyong Yuan}, \bibinfo{person}{Pan He},
  \bibinfo{person}{Qile Zhu}, {and} \bibinfo{person}{Xiaolin Li}.}
  \bibinfo{year}{2019}\natexlab{}.
\newblock \showarticletitle{Adversarial examples: Attacks and defenses for deep
  learning}.
\newblock \bibinfo{journal}{\emph{IEEE transactions on neural networks and
  learning systems}} \bibinfo{volume}{30}, \bibinfo{number}{9}
  (\bibinfo{year}{2019}), \bibinfo{pages}{2805--2824}.
\newblock


\bibitem[Zhou et~al\mbox{.}(2020)]%
        {zhou2020graph}
\bibfield{author}{\bibinfo{person}{Jie Zhou}, \bibinfo{person}{Ganqu Cui},
  \bibinfo{person}{Shengding Hu}, \bibinfo{person}{Zhengyan Zhang},
  \bibinfo{person}{Cheng Yang}, \bibinfo{person}{Zhiyuan Liu},
  \bibinfo{person}{Lifeng Wang}, \bibinfo{person}{Changcheng Li}, {and}
  \bibinfo{person}{Maosong Sun}.} \bibinfo{year}{2020}\natexlab{}.
\newblock \showarticletitle{Graph neural networks: A review of methods and
  applications}.
\newblock \bibinfo{journal}{\emph{AI Open}}  \bibinfo{volume}{1}
  (\bibinfo{year}{2020}), \bibinfo{pages}{57--81}.
\newblock


\bibitem[Z{\"u}gner et~al\mbox{.}(2018)]%
        {zugner2018adversarial}
\bibfield{author}{\bibinfo{person}{Daniel Z{\"u}gner}, \bibinfo{person}{Amir
  Akbarnejad}, {and} \bibinfo{person}{Stephan G{\"u}nnemann}.}
  \bibinfo{year}{2018}\natexlab{}.
\newblock \showarticletitle{Adversarial attacks on neural networks for graph
  data}. In \bibinfo{booktitle}{\emph{Proceedings of the 24th ACM SIGKDD
  International Conference on Knowledge Discovery \& Data Mining}}.
  \bibinfo{pages}{2847--2856}.
\newblock


\bibitem[Z{\"u}gner and G{\"u}nnemann(2019)]%
        {zugner2019adversarial}
\bibfield{author}{\bibinfo{person}{Daniel Z{\"u}gner} {and}
  \bibinfo{person}{Stephan G{\"u}nnemann}.} \bibinfo{year}{2019}\natexlab{}.
\newblock \showarticletitle{Adversarial Attacks on Graph Neural Networks via
  Meta Learning}. In \bibinfo{booktitle}{\emph{International Conference on
  Learning Representations}}.
\newblock


\end{thebibliography}

\end{document}